\documentclass[a4paper, 10pt]{article}
\usepackage[left=3cm, right=3cm, top = 3cm]{geometry}
\usepackage{chemformula} % Formula subscripts using \ch{}
\usepackage[T1]{fontenc} % Use modern font encodings
\usepackage{graphicx}
\usepackage{siunitx}
\usepackage{authblk}
\usepackage[utf8]{inputenc}
\usepackage[english]{babel}
\usepackage{amsmath}
\usepackage{todonotes}
\usepackage{textcomp}
\usepackage{braket}
\usepackage{gensymb}
\usepackage{caption}
\usepackage{csquotes}
\usepackage{cite}
\title{On--demand generation of entangled photon pairs in the telecom C--band for fiber--based quantum networks}

\author[1]{Katharina D. Zeuner\thanks{zeuner@kth.se}}
\author[1]{Klaus D. J\"ons\thanks{klausj@kth.se}}
\author[1]{Lucas Schweickert} 
\author[2]{Carl Reuterski\"old Hedlund} 
\author[2]{Carlos Nu\~nez Lobato} 
\author[1]{Thomas Lettner} 
\author[1]{Kai Wang} 
\author[1]{Samuel Gyger} 
\author[1]{Eva Sch\"oll} 
\author[1]{Stephan Steinhauer} 
\author[2]{Mattias Hammar} 
\author[1]{Val Zwiller} 
\affil[1]{Department of Applied Physics, Royal Institute of Technology, Albanova University Centre, Roslagstullsbacken 21, 106 91 Stockholm, Sweden}
\affil[2]{Department of Electrical Engineering, Royal Institute of Technology, Electrum 229, 164 40 Kista, Sweden}

\begin{document}

\maketitle

\begin{abstract}
%Intro in 150 words...
On--demand sources of entangled photons for the transmission of quantum information in the telecom C--band are required to realize fiber--based quantum networks. So far, non--deterministic sources of quantum states of light were used for long distance entanglement distribution in this lowest loss wavelength range. However, they are fundamentally limited in either efficiency or security due to their Poissonian emission statistics. Here, we show on--demand generation of entangled photon pairs in the telecom C-band by an InAs/GaAs semiconductor quantum dot. Using a robust phonon--assisted excitation scheme we measure a concurrence of \SI{91.4}{\%} and a fidelity of \SI{95.2}{\%} to $\Phi^+$. On--demand generation of polarization entangled photons will enable secure quantum communication in fiber--based networks. Furthermore, applying this excitation scheme to several remote quantum dots tuned into resonance will enable first on--demand entanglement distribution over large distances for scalable real--life quantum applications. 

\end{abstract}
%\maketitle
\section*{Introduction}
Quantum communication will yield unconditional security permitted by quantum cryptography protocols~\cite{Ekert, Bennett1992} via the transfer of single and entangled photons, as well as connect remote quantum computers, e.g. for cloud quantum computing~\cite{Devitt2016}. Long distance distribution of photonic qubits requires quantum sources of light emitting at telecom wavelengths, ideally in the telecom C--band (\SI{1530}{\nano\metre} - \SI{1565}{\nano\metre}) to benefit from the lowest absorption loss in optical fibers. Historically, photons generated by spontaneous processes, e.g. parametric down conversion or four--wave mixing have dominated the field of fiber--based entanglement distribution~\cite{Li2005, Hubel2007, Treiber2009,  Dynes2009, Wengerowsky2019}, but the random photon flux and timing of the emission process severely limits implementation. In recent years semiconductor quantum dots have emerged as strong competitors due to their promise of deterministic qubit generation, based on their preeminent emission of on--demand single photons~\cite{Schweickert2018, Hanschke2018} and entangled photon pairs~\cite{Muller2014, Winik2017, Huber2018, Chen2018, Liu2019, Wang2019}. These outstanding properties hinge on the emission of photons via the radiative biexciton exciton cascade, emitting polarization entangled photons~\cite{Benson2000}. In the presence of asymmetry, the excitonic states exhibit a finestructure splitting resulting in the time evolving two--photon Bell state generated by the radiative cascade~\cite{Ward2014}:
\begin{equation}
    \ket{\Psi_{\text{ev}}}=\frac{1}{\sqrt{2}}\left(\ket{\text{HH}}+e^{\frac{i\delta\text{t}}{\hbar}}\ket{\text{VV}}\right).
\end{equation} With appropriate time resolution the oscillating entangled state can be resolved and the measured degree of entanglement is unaltered~\cite{Winik2017, Fognini2019}.
Alongside suitable time resolution, a resonant excitation scheme is required for the generation of unprecedented degrees of entanglement~\cite{Fognini2019}, which is provided by two--photon resonant excitation~\cite{Stufler2006}. So far, two--photon resonant excitation has been used to demonstrate background--free emission of single photons~\cite{Schweickert2018, Hanschke2018}, high degrees of entanglement with near--unity fidelity and concurrence~\cite{Huber2018}, quantum teleportation~\cite{Reindl2018} and entanglement swapping~\cite{Basset2019, Zopf2019}, however all outside the telecom range. While this excitation scheme has produced impactful results in these experiments, practical quantum networks relying on the interaction of multiple remote sources require excitation techniques that function for a variety of quantum dots and independently of environmental fluctuations to provide a building block for a scalable and robust quantum architecture. Besides requiring very quantum dot specific excitation conditions, small fluctuations in excitation laser power or wavelength would cause diminished state occupation for the quantum dots, rendering the two--photon resonant scheme impractical for large networks.
A more robust and universal scheme compared to pure two--photon excitation is phonon--assisted two--photon excitation~\cite{Glassl2013}. It enables to prepare the state with close to maximum probability for a much broader range of excitation powers and wavelengths, while keeping consistently high levels of fidelities and indistinguishability of the prepared photons~\cite{Reindl2017}, making the scheme highly relevant for future applications that require several remote emitters at the same wavelength such as quantum teleportation or entanglement swapping.
Entangled photon pairs at telecom wavelengths have been demonstrated with quantum dots in the O--band~\cite{Ward2014}, S--band~\cite{Muller2018} and C--band~\cite{Olbrich2017}, however only under continuous, non--resonant excitation. \newline
\section*{Results}
\subsection*{Telecom entanglement setup} %add from methods
Here, we present on-demand generation of entangled photon pairs from epitaxially grown InAs quantum dots with a metamorphic--buffer layer on a GaAs substrate enabling emission in the telecom C--band (see methods section)~\cite{Paul2017}. This industry grade growth technique yields scalable sources, while avoiding short comings of InP--based emitters. 
A schematic illustration of the setup is presented in Fig. \ref{fig:setup}, showing the excitation laser (a), cryostat and sample (b), single photon detection (d), filtering with entanglement analysis (e) and spectroscopy setup (f) (for more details see methods). 
The sample is placed in a closed--cycle cryostat and cooled to \SI{10}{K} (Fig.~\ref{fig:setup} (b)). To excite the sample we use a tunable pulsed laser that generates \SI{2}{\pico\second} pulses with a repetition rate of \SI{80}{MHz}. A pulse slicer is used to adjust the bandwidth of the laser pulses between 2 and \SI{70}{\pico\second} (Fig.~\ref{fig:setup} (a)). After excitation the emitted quantum dot photons are collected with a 0.8\,NA objective and then sent to our spectroscopy (Fig.~\ref{fig:setup} (b) and (f)) and entanglement analysis setup (Fig.~\ref{fig:setup} (e)). Tunable notch filters (F) with \SI{0.7}{\nano\meter} spectral bandwidth can be used to either block the excitation laser (F1 and F2) or to reflect a selected quantum dot transition and separate XX from X and from remaining laser light (F3 and F4). F3 and F4 can be tuned such that the entire quantum dot spectrum can be sent to a spectrometer equipped with an InGaAs array for spectral analysis (typical resolution is \SI{25}{\micro\eV}). If tuned to the exciton and biexciton wavelengths, F3 and F4 deflect the quantum dot photon towards standard telecom single mode fibers. The fibers carrying the XX or X photons can be connected to C1 for spectral analysis. A set of waveplates, as well as a polarizer in front of each fiber are used to set the polarization basis for the quantum state tomography measurements. The fiber--coupled quantum dot photons are then connected to superconducting nanowire single photon detectors with a time resolution of approximately \SI{20}{\pico\second} and efficiencies of \SI{15}{\%} and \SI{25}{\%} measured at a dark count level of \SI{30}{\per\second} (Fig.~\ref{fig:setup} (d)).
\subsection*{Photoluminescence measurements}
\begin{figure}
\includegraphics[width=\columnwidth]{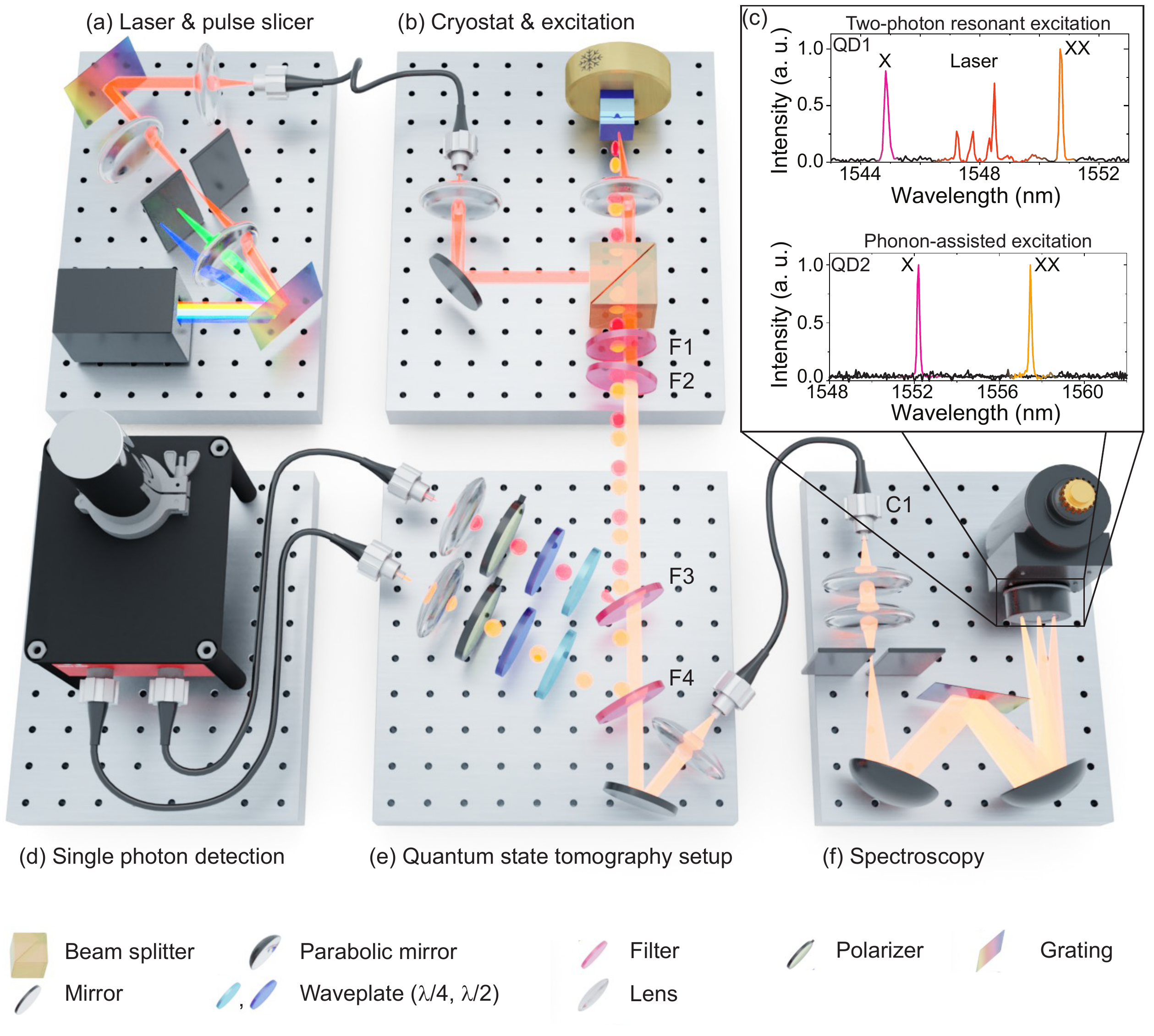}
\caption{\label{fig:setup} Telecom entanglement setup consisting of (a) laser excitation and pulse shaper, (b) cryogenically cooled InAs/GaAs quantum dot sample with excitation setup, (d) superconducting nanowire single photon detectors for time resolved measurements, (e) filtering and quantum state tomography setup and (f) spectroscopy setup. (c) Quantum dot spectra for two--photon resonant excitation (top) and phonon--assisted two--photon resonant excitation (bottom). The TPE spectrum is recorded after filtering only with F1 and F2, the spectra under phonon--assisted excitation are measured after the exciton (X) is reflected from F3 and the biexciton (XX) from F4, respectively.}
\end{figure}

In the top part of Fig. \ref{fig:setup} (c) we show a quantum dot spectrum in the telecom C-band recorded under two--photon resonant excitation (QD1). Excitonic and biexcitonic emission are visible in the spectrum at emission wavelengths of \SI{1544.8}{\nano\meter} and \SI{1550.7}{\nano\meter}, respectively. In between the two emission lines, scattered laser light is visible (red). 
In the bottom part of Fig. \ref{fig:setup} (c) we show the spectra for exciton and biexciton of QD2 using the phonon--assisted two--photon excitation scheme. The spectra are recorded through optical fibers after the tunable filters F3 and F4 in addition to tunable fiber--based filters (not shown) tuned to either the emission wavelength of the exciton or the biexciton. As clearly visible in the spectrum, the remaining contribution of the excitation laser after our careful filtering setup is negligible. 
%%%%%%%%%%%%%%%%%%%%%%%%%%%%%%%%%%%%%%%%%%%%%%Figure1%%%%%%%%%%%%%%%%%%%%%%%%%%%%%%%%%%%%%%%%%%%%%%%%%%%%%%%%%%%%%%%%
%%%%%%%%%%%%%%%%%%%%%%%%%%%%%%%%%%%%%%%%%%%%%%%%%%%%%%%%%%%%%%%%%%%%%%%%%%%%%%%%%%%%%%%%%%%%%%%%%%%%%%%%%%%%%%%%%
\subsection*{Two-photon excitation schemes}
\begin{figure}
\includegraphics[width=\columnwidth]{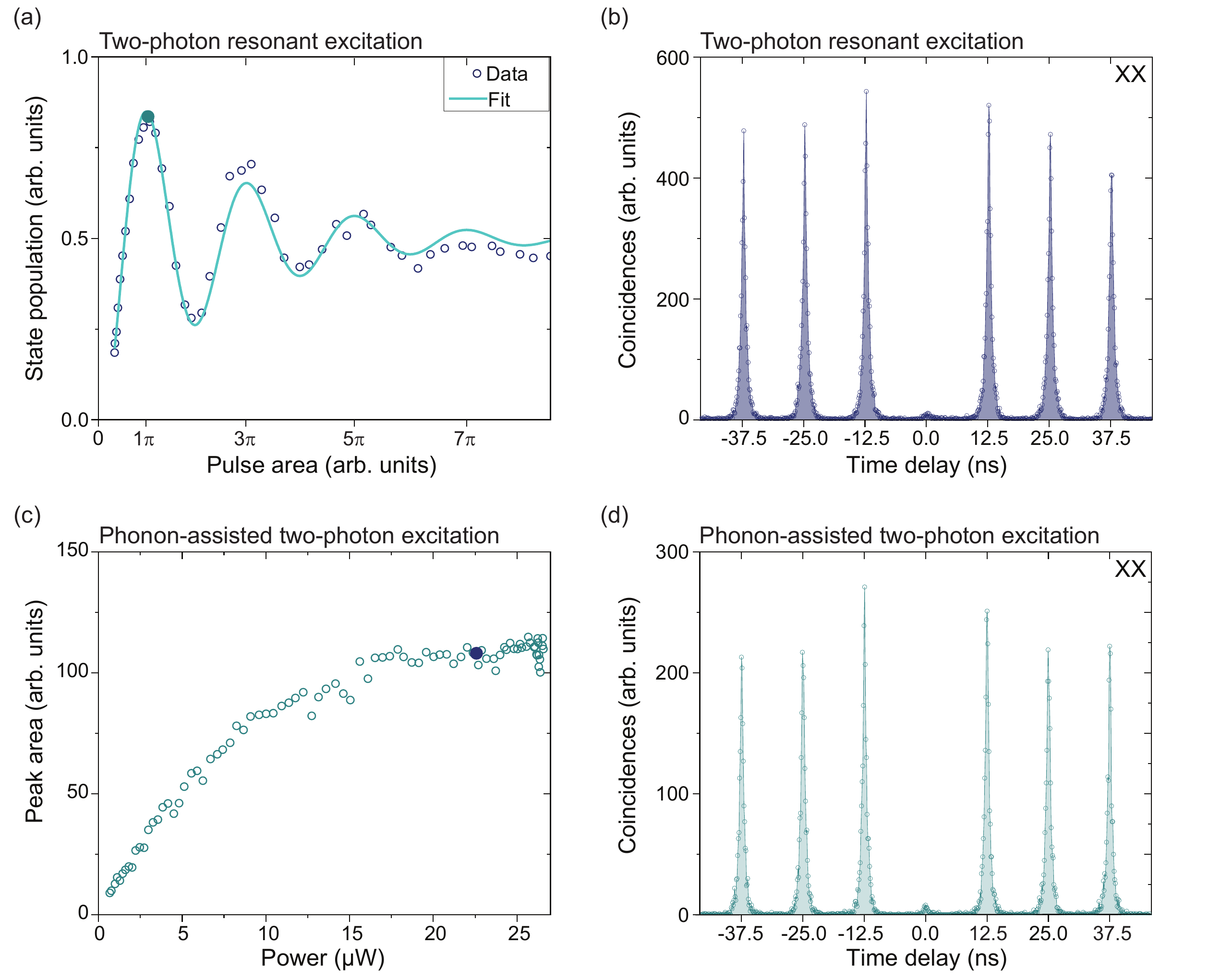}
\caption{\label{fig:TPE} Two--photon resonant excitation in comparison with phonon--assisted excitation. (a) Excitation power--dependent peak area of the exciton state of QD1 showing Rabi oscillations up to $7\,\pi$. The filled dot represents the pulse area used to measure the autocorrelation. (b) Autocorrelation measurement of the biexciton state under $\pi$--pulse excitation. (c) Excitation power--dependent peak area of the exciton state of QD2 showing plateau--like saturation behavior for higher excitation powers. The filled dot represents the pulse area used to perform all following measurements. (d) Autocorrelation measurement of the biexciton state.}
\end{figure}

Coherent control of the two--level system of QD1 is demonstrated via the integrated peak area as a function of excitation power shown in Fig. \ref{fig:TPE} (a) for the exciton with Rabi oscillations up to $7\,\pi$. With a fit to the data we can extract the population in the $\pi$--pulse of \SI[parse-numbers = false]{82.1\pm1.2}{\percent}. To stabilize the charge environment, we add approximately \SI{100}{\nano\watt} of a continuous--wave HeNe laser to our pulsed excitation laser (see supplementary information). To determine the single photon purity, we perform a second--order autocorrelation measurement of the biexciton under two--photon resonant $\pi$--pulse excitation, which is shown in Fig. \ref{fig:TPE} (b), yielding $g^{(2)}(0)=0.043\pm0.004$.
Polarization--dependent photoluminescence measurements reveal a finestructure splitting of \SI[parse-numbers = false]{24.6 \pm 7.1}{\micro\eV} for this particular quantum dot (QD1). This results in a precession of the eigenstate with a period of $\text{T}=\frac{\text{h}}{\delta}$=\SI{168}{\pico\second}, making this quantum dot unsuitable for entanglement measurements with our setup time resolution (\SI{30}{\pico\second}).
%%%%%%%%%%%%%%%%%%%%%%%%%%%%%%%%%%%%%%%%%%%%%Figure2%%%%%%%%%%%%%%%%%%%%%%%%%%%%%%%%%%%%%%%%%%%%%%%%%%%%%%%%%%%%

%%%%%%%%%%%%%%%%%%%%%%%%%%%%%%%%%%%%%%%%%%%%%%%%%%%%%%%%%%%%%%%%%%%%%%%%%%%%%%%%%%%%%%%%%%%%%%%%%%%%%%%%%%%%%%%%
A power dependent measurement of QD2 under phonon-assisted resonant excitation yields the curve shown in Fig.~\ref{fig:TPE} (c), similar to the one previously reported in Ref.~\cite{Reindl2017} at \SI{780}{\nano\meter}. All following measurements are taken with excitation powers well within the plateau at the end of the power--series curve. We also conduct a second--order autocorrelation measurement for QD2 which yields $g^{(2)}(0)=0.038\pm0.005$, similar to the case of the pure two--photon resonant excitation used for QD1. From the time--tagged autocorrelation measurement, a biexciton lifetime of $445.66 \pm 4.25\,\si{\pico\second}$ is extracted (see also supplementary material). 
\subsection*{Generation of highly entangled photons}
To demonstrate that we can extract entangled photon pairs under phonon--assisted two--photon resonant excitation, we perform quantum state tomography on the two photon state resulting from the XX-X-cascade by recording time--tagged histograms in all 36 different polarization bases~\cite{James}, which we evaluate with our extensible time tag analyzising software~\cite{ETA}. 
%%%%%%%%%%%%%%%%%%%%%%%%%%%%%%%%%%%%%%%%%%%%%%%%%Figure3%%%%%%%%%%%%%%%%%%%%%%%%%%%%%%%%%%%%%%%%%%%%%%%%%%%%%%
\begin{figure}
\includegraphics[width=\columnwidth]{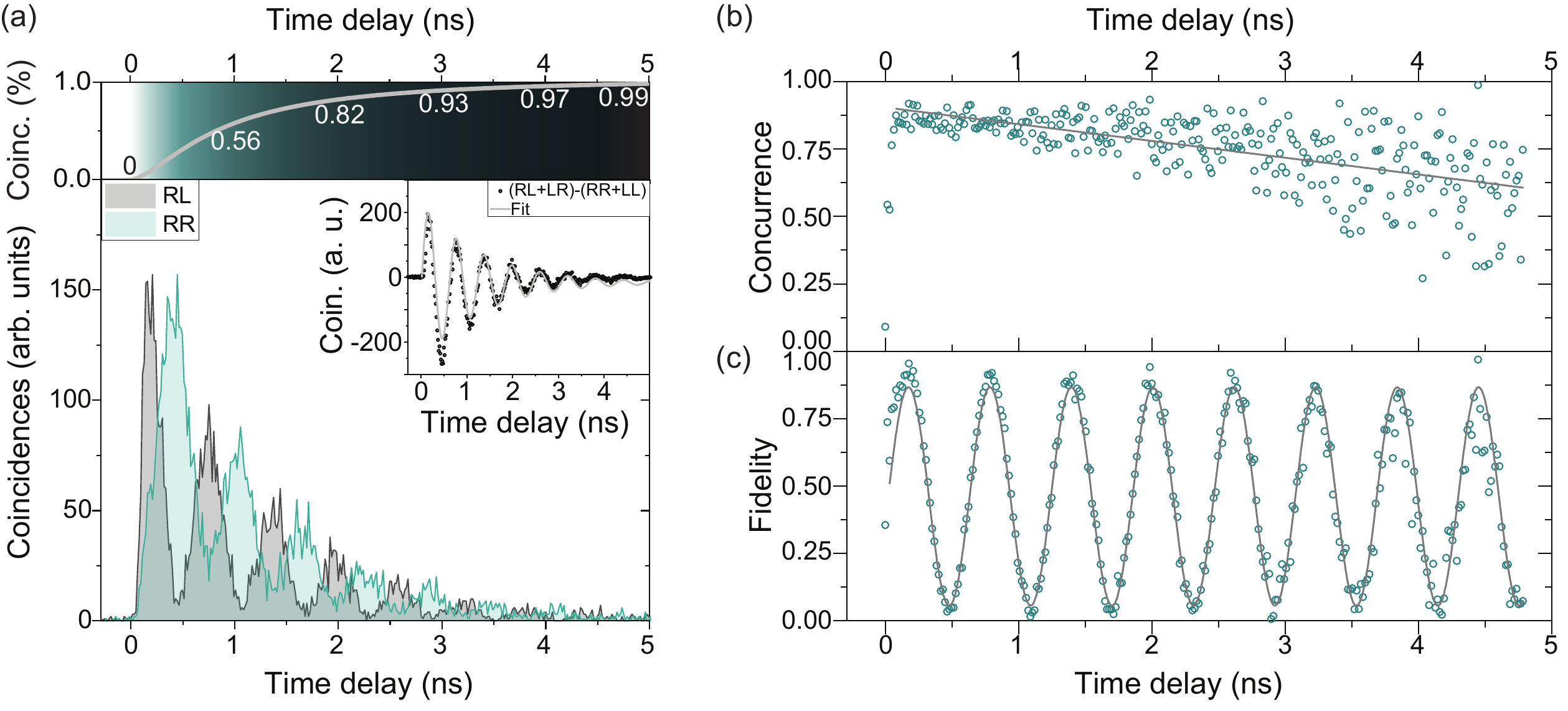}
\caption{\label{fig:Entanglement} Quantum state tomography in the telecom C--band. (a) Top panel: amount of total coincidences as a function of time. Center peaks of two coincidence measurements recorded in the circular basis (RR and RL) that are showing oscillations due to the finestructure splitting of QD2. Inset: quantum oscillations between the two Bell states $\Phi^+$ and $\Phi^-$. (b) Concurrence reconstructed from the quantum state tomography measurements. The green open circles correspond to data, the gray solid line corresponds to a linear fit to the data. The maximum concurrence is \SI{91.4}{\%}$\pm$\SI{3.8}{\%} for a time delay of \SI{176}{\pico\second}. (c) Fidelity to $\Phi^+$ as a function of time, showing oscillations due to the finestructure splitting of QD2. The green open circles correspond to data, the gray solid line corresponds to a sine fit to the data.}
\end{figure}
\begin{figure}
\includegraphics[width=\columnwidth]{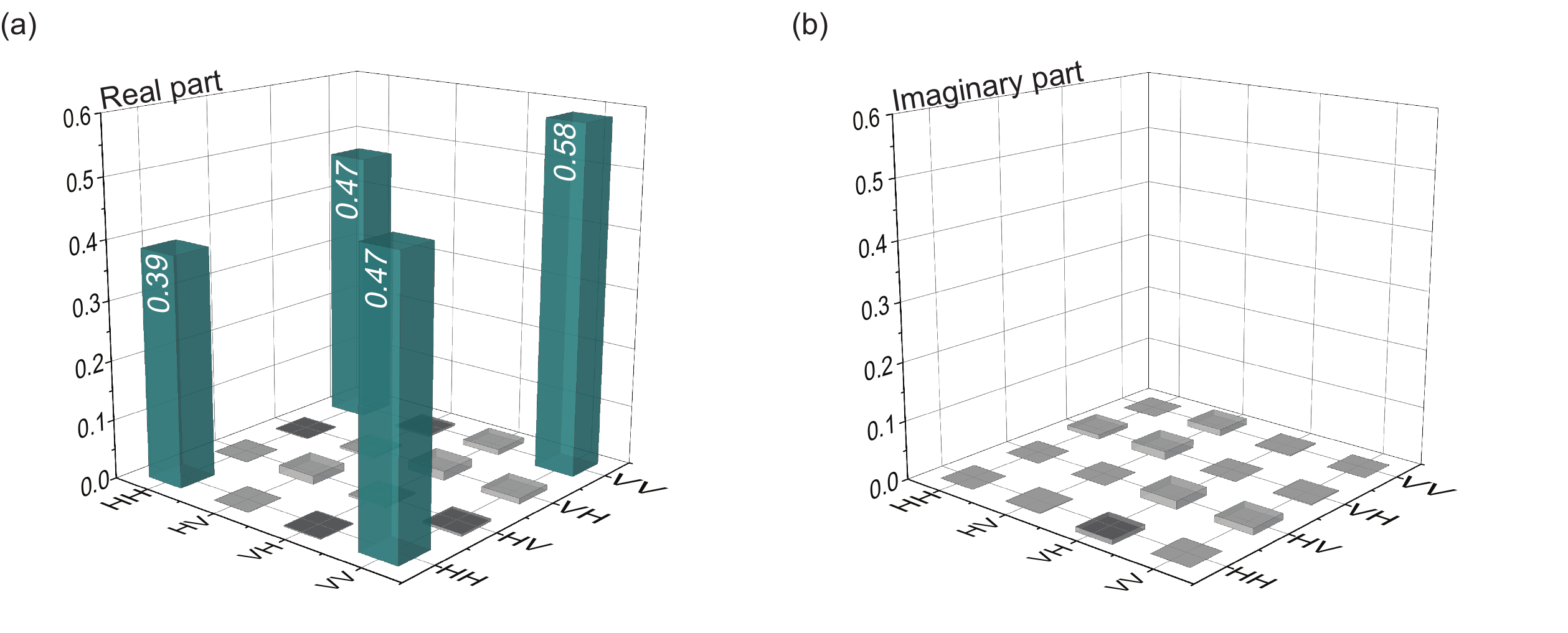}
\caption{\label{fig:DM} Density matrix. (a) Real part of the density matrix reconstructed for a time delay of \SI{176}{\pico\second} and a bin width of \SI{16}{\pico\second}. (b) Corresponding imaginary part of the density matrix for (a).} 
\end{figure}
%%%%%%%%%%%%%%%%%%%%%%%%%%%%%%%%%%%%%%%%%%%%%%%%%%%%%%%%%%%%%%%%%%%%%%%%%%%%%%%%%%%%%%%%%%%%%%%%%%%%%%%%%%%%%%%%
The center peak in the coincidence histogram shows oscillations (Fig.~\ref{fig:Entanglement} (a)) due to the evolving polarization state of the exciton in the circular basis. The emitted photon state temporally oscillates between the two Bell states $\Phi^+=1\sqrt{2}(\ket{\text{HH}}+\ket{\text{VV}})=1\sqrt{2}(\ket{\text{RL}}+\ket{\text{LR}})$ and $\Phi^-=1\sqrt{2}(\ket{\text{HH}}-\ket{\text{VV}})=1\sqrt{2}(\ket{\text{RR}}+\ket{\text{LL}})$, which is shown in the inset of Fig.~\ref{fig:Entanglement} (a). From the oscillations we can extract a value of the finestructure splitting of \SI{4.11}{\micro\eV}, yielding a more precise value due to higher resolution in the time domain compared to spectroscopy. To analyze the degree of entanglement, we extract the concurrence from the quantum state tomography data using a modified version of the quantum state tomography code of Ref.~\cite{TjeerdFokkensAndreasFognini}. Within the full displayed range of \SI{5}{\nano\second}, corresponding to the period of time in which \SI{99}{\%} of the correlations have taken place (see top panel of Fig.~\ref{fig:Entanglement} (a)), the concurrence is well above $50\,\%$, shown in Fig.~\ref{fig:Entanglement} (b). A concurrence larger than 0 confirms the presence of entangled photons. Furthermore, we extract a maximum concurrence of $91.4\pm3.8\,\%$ as a result of our excitation scheme combined with the high time resolution of the detectors~\cite{Fognini2019}. Figure~\ref{fig:Entanglement} (c) shows the fidelity to the state $\Phi^+=\frac{1}{\sqrt{2}}(\ket{\text{HH}}+\ket{\text{VV}})$ as a function of time. Also here the characteristic oscillations due to the finestructure splitting of the quantum dot can be observed, making the emitted state oscillate between $\Phi^+$ and $\Phi^-$. The maximum fidelity to $\Phi^+$ is $95.2\pm1.1\,\%$ for a time delay of \SI{176}{\pico\second}, corresponding to a quantum bit error rate as low as \SI{3.2}{\%}.
Finally, we reconstruct the density matrix from the quantum state tomography measurement, which is presented in Fig.~\ref{fig:DM} for a time delay of \SI{176}{\pico\second} and a bin width of \SI{16}{\pico\second} after it has undergone a coordinate transformation to compensate for birefringence caused by the setup (see methods section and supplementary).  The real part of the density matrix (Fig.~\ref{fig:DM} (a)) exhibits dominant outer diagonal elements, while all other elements of the matrix are strongly suppressed, with negligible imaginary part (Fig.~\ref{fig:DM} (b)). This further highlights the unrivaled quality of the entangled photons created from our source via the phonon--assisted excitation scheme. 
\section*{Discussion}
We have demonstrated on--demand emission of polarization entangled photon pairs from a InAs/GaAs quantum dot in the telecom C--band. Besides the small finestructure splitting, we are able to measure a maximum concurrence of up to $91.4\pm3.8\,\%$ and a concurrence larger than \SI{50}{\%} within a time window of \SI{5}{\nano\second}. An unprecedented level of fidelity in the telecom C--band to $\Phi^+$ of $95.2\pm1.1\,\%$ was extracted without correcting for detector dark counts, background photons or the non--zero $\text{g}^{(2)}(0)$-value. The high quality entanglement that we are generating with our source is based on the state preparation via the phonon--assisted two--photon excitation scheme, combined with a good enough time resolution compared to the fine structure splitting of our quantum dot. The biexciton lifetime extracted under phonon--assisted excitation is significantly shorter than previously predicted by non--resonant excitation methods and demonstrates that these InAs/GaAs quantum dots could be operated at rates above the usual \SI{80}{\mega\hertz}, allowing high quantum key rates. The industry grade growth technique of our quantum dots will make entangled photon emitters in the telecom C--band widely available and is, thus, providing feasible sources for deployment in fiber—based quantum networks. The high level of concurrence in combination with the resilient phonon--assisted excitation scheme has strong potential for any application relying on remote sources of entangled photons. Furthermore, the on--demand generation of entangled photons opens up possibilities to transmit quantum secure keys efficiently over long distances, marking a step towards practical applications of quantum dots in fiber--based quantum networks.
\section*{Methods}
%\subsection*{Setup}
 %
\subsection*{Sample - Growth}
The sample was grown by metal--organic vapor--phase epitaxy (MOVPE) on Si--doped GaAs (001)--oriented substrates in an Aixtron 200/4 low-pressure (\SI{100}{\milli\bar)} horizontal reactor with $\text{H}_2$ as carrier gas and trimethylgallium (TMGa), trimethylaluminium (TMAl), trimethyl-indium (TMIn), and arsine ($\text{AsH}_3$) as precursors. The epitaxial layer structure is given Table 1. The distributed Bragg reflector (DBR) and compositionally graded InGaAs metamorphic buffer layer (MMBL) were first grown at \SI{670}{\celsius} (calibrated wafer surface temperature) after which the growth was stopped and the temperature was reduced to \SI{515}{\celsius} for quantum dot growth. Next, a \SI{10}{\second} ripening step was used and the low--temperature part of the capping layer was grown. Finally, the temperature was increased to \SI{670}{\celsius} and the structure was completed with the high--temperature part of the capping layer. A three--lambda cavity is formed between the DBR and the semiconductor--air interface with the MMBL and capping layer thicknesses chosen to optimize the extraction efficiency. The lattice relaxation of the MMBL layer allows for the growth of large QDs with an emission wavelength of around \SI{1550}{\nano\meter}, which is significantly longer than what can be obtained from the coherent growth on the GaAs substrate (typically <\SI{1300}{\nano\meter})~\cite{Paul2017}. In our previous work, using similar growth conditions, we estimated the QD density to be in the \SI{1e7}{\per\square\centi\meter} range~\cite{Zeuner2018}.
\begin{table}[htbp]
\centering
\begin{tabular}{l|llll}
Layer & Material & Thickness & Growth temperature & Comment\\\hline
Substrate & GaAs & \SI{350}{\micro\meter} & &\\
DBR (x19.5) & AlAs/GaAs & \SI{134.4}{\nano\meter}/\SI{114.6}{\nano\meter} & \SI{670}{\celsius} & \\
MMBL & $\text{In}_\text{x}\text{Ga}_{1-\text{x}}\text{As}$ & \SI{1150}{\nano\meter} & \SI{670}{\celsius} & x=0.015-0.4\\
QD & InAs &  & \SI{515}{\celsius} & \\
Low T Cap & $\text{In}_\text{x}\text{Ga}_{1-\text{x}}\text{As}$ & \SI{10}{\nano\meter} & \SI{515}{\celsius} & x=0.3\\
High T Cap & $\text{In}_\text{x}\text{Ga}_{1-\text{x}}\text{As}$ & \SI{195}{\nano\meter} & \SI{670}{\celsius} & x=0.3\\\hline
\end{tabular}
\caption{Details of used layer structure and growth parameters for quantum dot emission in the telecom C--band}
\label{tab:growth}
\end{table}
\subsection*{Reconstruction of the quantum state}
The density matrix shown in Fig. \ref{fig:DM} is corrected for birefringence introduced by the capping layer of the sample, the cryostat and the excitation part of the setup (see Fig. \ref{fig:setup} (b)). This performs a coordinate transformation of the photons that are originally emitted in the standard HV--basis, to the birefringent $\Tilde{\text{H}}\Tilde{\text{V}}$--basis. A virtual waveplate is introduced in order to transform back from the birefringent $\Tilde{\text{H}}\Tilde{\text{V}}$-coordinate system to the HV--basis in which the polarization analysis is performed. The applied transformation is preserving the orthogonality of the polarization basis such that the absolute values of e.g. the fidelity of a state in a given basis compared to a maximally entangled state are not changed. The maximally entangled state in the birefringent $\Tilde{\text{H}}\Tilde{\text{V}}$--basis, as well as the closest entangled state in the HV--basis are given in the supplementary.

\section*{Acknowledgements}
This project has received funding from the European Union's Horizon 2020 research and innovation program under grant agreement No. 820423 (S2QUIP). K.D.J. acknowledges funding from the Swedish Research Council (VR) via the starting grant HyQRep (Ref.: 2018-04812) and the The Göran Gustafsson Foundation (SweTeQ). M. H. acknowledges funding from the Swedish Research Council (VR, grant No. 2016-03388). V.Z. acknowledges funding by the European Research Council under the Grant Agreement No. 307687 (NaQuOp), the Knut and Alice Wallenberg Foundation (KAW, ”Quantum sensors”) and the Swedish Research Council (VR, grant No. 638-2013-7152 and grant No. 2018-04251). The Quantum Nano Photonics group at KTH acknowledges the continuous support by the companies APE Angewandte Physik und Elektronik GmbH on their picoEmerald system and Single Quantum BV on their detector system. K.D.Z. and K.D.J. acknowledge fruitful discussions with M. Rota and R. Trotta. K.D.J. acknowledges fruitful discussions with Matthias Paul. T.L. acknowledges fruitful discussions with Marijn Versteegh.

\section*{Author contributions}
The sample was grown by C.R.H., C.N.L., and M.H. Sample characterization was performed by K.D.Z, S.S., and K.W. with help from C.N.L. K.D.Z., L.S., and K.D.J. built the setup with help from T.L. The experiment was performed by K.D.Z, and L.S, with help from K.D.J., S.G., and E.S. The data analysis was performed by K.D.Z., L.S, K.D.J., and T.L. The paper was written by K.D.Z and V.Z. with input from all authors. The project was conceived by K.D.Z., K.D.J., and V.Z. and supervised by K.D.J., and V.Z.

\section*{Competing interests}
The authors declare no competing interests.

%\bibliography{telecom_entanglement.bib}{}
%\bibliographystyle{ieeetr}

\end{document}